\documentclass[prb,aps,twocolumn,showpacs]{revtex4}
\usepackage{graphicx,color,latexsym}
\usepackage{dcolumn}
\usepackage{amsmath,amssymb,epsf,bm}
\begin{document}
\title{Non-exponential long-range interaction of magnetic impurities in spin-orbit coupled superconductors.}
\author{A.~G. Mal'shukov}
\affiliation{Institute of Spectroscopy, Russian Academy of Sciences, Troitsk, Moscow, 108840, Russia}
\affiliation{Moscow Institute of Physics and Technology, Institutsky per.9, Dolgoprudny, 141700 Russia}
\begin{abstract}
The Ruderman-Kittel-Kasuya-Yosida (RKKY) interaction of magnetic impurities  in a superconductor exponentially decreases when the distance $r$ between them is larger than the superconductor's coherence length, because this interaction is mediated by quasiparticles, which have a gap in their energy spectrum. At the same time, the spin-singlet superconducting condensate was always assumed to stay neutral to magnetic impurities. Due to a spin-orbit coupling (SOC), however, Cooper pairs gain an admixture of spin-triplet correlated states, which provide for a link between impurity spins and an s-wave condensate. It is shown that perturbations of its phase mediate the $1/r^2$ interaction of these spins in two-dimensional (2D)  systems. This effect is considered within two models: of a clean 2D s-wave superconductor with the strong Rashba SOC and of a bilayer system which combines a 2D Rashba coupled electron gas and an s-wave superconducting film. The predicted long-range interaction can have a strong effect on spin orders in superconductor-magnetic impurity systems that are expected to host Majorana fermions.
\end{abstract}
\maketitle

\section{Introduction}

RKKY interaction of localized spins in metals \cite{Ruderman,Kasuya,Yosida} is carried by spin excitations. A localized spin, by coupling to spins of conduction electrons through the exchange interaction, creates one-particle spin excitations close to the Fermi surface. They, in turn, exert influence upon the spin of an impurity placed at some distance. This results in an effective exchange interaction between impurity spins. Recently, the RKKY interaction has attracted a renewed attention in connection with the search of Majorana quasiparticles  localized at 1D and 2D magnetic impurity systems on the surface of superconductors \cite{Nadj,Pientka,Christensen,Klinovaya,Menard,Pawlak,Li}. In this context the presence of a strong SOC has been found important for the formation of the topological  phase in superconductors and superconductor proximity systems  \cite{Li,Ptok,Hui,Nadj2,Brydon,Heimes}. For example, in spin-orbit coupled superconductors the RKKY interaction contains the Dzyaloshinskii-Moryia term  \cite{Heimes,Zyuzin,Qin} which favors a spiral spin order in impurity spin chains.

In superconductors having a gap in the quasiparticle spectrum the RKKY interaction exponentially decreases with the distance between impurities \cite{Anderson}. Such a behavior was also found in spin orbit coupled superconductor proximity systems \cite{Zyuzin}. In both cases the interaction range is restricted by the Cooper pair coherence length. This is because at low energies only evanescent quasiparticle waves can propagate in a gapped superconductor. While such virtual quasiparticles mediate interaction of impurity spins, the condensate of singlet Cooper pairs does not participate in this process. This is obviously true, if a SOC is absent. At the same time, in the presence of  a strong SOC the role of the singlet condensate should be revised.  Below, the influence of the superconducting condensate on the  interaction of localized spins will be studied for a helical  2D system, that is a system with the strong Rashba SOC \cite{Rashba}. Such a system may be a topological s-wave superconductor on the surface of a  3D topological insulator (TI), or a non-topological superconductor with strong enough Rashba SOC. It will be shown that in such systems the SOC provides a coupling of impurity spins to the condensate. Therefore, the condensate gains the ability to mediate the interaction between magnetic impurities. Since this interaction propagates through the condensate, it is important to distinguish between intrinsic superconductors and proximity induced superconducting systems. In the latter case the problem is more complicated, because magnetic impurities may be placed in the  normal spin-orbit coupled system, while interaction between them is mediated by the superconductor. However, it will be shown that in both cases at large distances the interaction decreases algebraically, as $1/r^2$. Basically, such a long-range effect originates from invariance of the system with respect to a spatially uniform shift  of the  condensate phase.

Two models will be considered. The first one represents a 2D clean helical superconductor with the isotropic electron-electron BCS interaction. In the second model a superconducting film makes a contact with the surface of a 3D TI and a pair of magnetic impurities are placed onto this surface. Both, the  film and TI surface are assumed to be dirty systems. The spin dynamic of impurities will be ignored. Therefore, their spins are static classical spins $S\gg \hbar$. Also, the exchange interaction between  spins of impurities  and conduction electrons will be treated perturbatively by taking into account only its leading orders. Therefore, a possibility for formation of Yu-Shiba-Rusinov states inside the superconductor's gap is excluded in such an approach, although these states can play an important role in short-range interactions of magnetic impurity \cite{Yao}.

The article is organized in the following way. In Sec.II the condensate mediated interaction of magnetic impurities is calculated for a spin-orbit coupled 2D superconductor. In Sec.III the influence of the calculated long-range interaction is analyzed for various spin arrays. In Sec. IV the condensate mediated interaction of impurity spins is considered for a bilayer system of a superconductor  and a 2D Rashba normal metal where superconducting correlations are induced due to the proximity effect. A summary of results is presented in Sec.V.

\section{The condensate mediated interaction of spins in a two-dimensional superconductor}

Let us consider a two-dimensional superconductor with an isotropic attractive electron-electron interaction and the strong Rashba SOC. A pair of magnetic impurities is placed at the points $\mathbf{r}_1$ and $\mathbf{r}_2$. Their spins interact with conduction electrons according to the exchange Hamiltonian $H_{\mathrm{int}}=\sum_{\nu} J_{ij}S^i_{\nu}\sigma^j_{\nu}$, where $\nu=1,2$ and $\sigma^j_{\nu}=\psi^{\dag}(\mathbf{r}_{\nu})\sigma^j\psi(\mathbf{r}_{\nu})$ , with $\sigma^j$ denoting Pauli matrices ($j=x,y,z$). The field operators $\psi(\mathbf{r}_{\nu})$ are vectors defined in the Nambu basis as $\psi=(\psi_{\uparrow},\psi_{\downarrow},\psi^{\dag}_{\downarrow},-\psi^{\dag}_{\uparrow})$. The interaction between impurity spins is given by the second-order correction to the free energy due to $H_{\mathrm{int}}$  \cite{AGD}
\begin{equation}\label{U}
U_{12}=-\int_0^{\beta}d\tau Z^{j}_1Z^{m}_2\langle\mathrm{Tr}[\mathrm{T}\left(\sigma^j_1(\tau)\sigma^m_2(0)\right)]\rangle\,,
\end{equation}
where $\beta=1/T$ ($T$ is the temperature), $Z^{j}_{1(2)}=J_{ij}S^i_{1(2)}$, and the angular brackets denote the thermodynamic average over unperturbed states.  The trace in (\ref{U}) is taken over spin and Nambu variables and $\mathrm{T}$ is the Matsubara time ordering operator. We set $\hbar=1$ and the Boltsmann constant $k_B=1$. The  unperturbed Hamiltonian is given by the sum $H=H_0+V$ of the one-particle Hamiltonian $H_0$ and the two-particle attractive interaction $V$. $H_0=\sum_{\mathbf{k}}\psi^{\dag}_{\mathbf{k}}\mathcal{H}_{0\mathbf{k}}\psi_{\mathbf{k}}$ represents a spin-orbit coupled 2D electron gas, where $\psi_{\mathbf{k}}$ are the electron field operators in the wave-vector representation and $\mathcal{H}_{\mathbf{k}}$ is given by
\begin{equation}\label{H0}
\mathcal{H}_{0\mathbf{k}}=\tau_3(\epsilon_{\mathbf{k}}-\mu)+\tau_3\mathbf{h}_{\mathbf{k}}\bm{\sigma}
\end{equation}
where $\epsilon_{\mathbf{k}}=k^2/2m$ and $\mu$ are the electron band energy and the chemical potential, respectively. The Pauli matrices $\tau_i$, $i=0,1,2,3$, operate in the Nambu space, where $\tau_0$ is the unit matrix. The second term in $\mathcal{H}_0$ represents the Rashba SOC whose spin-orbit field is given by $h^x_{\mathbf{k}}=-\alpha k_y$ and $h^y_{\mathbf{k}}=\alpha k_x$ \cite{Rashba}. This interaction results in a splitting of the conduction band into two helical bands with opposite helicities $\gamma=\pm 1$, so that the average electron's spins in these bands are parallel to $\gamma \mathbf{h}_{\mathbf{k}}$. In each of the bands electrons have the same Fermi velocity $v_F=\sqrt{2\mu/m +\alpha^2}$ and different state densities $N_{F\gamma }=(m/2\pi)(1-\gamma \alpha/v_F)$.

The two-particle interaction $V$  is assumed to be independent on wave-vectors of interacting particles, except for the high-energy cutoff $\omega_c$ near the Fermi surface. Therefore, it is given by
\begin{equation}\label{V}
V= \frac{g}{2} \sum_{\mathbf{k,k^{\prime},q}}(\psi^{\dag}_{\mathbf{k}}\tau_3\psi_{\mathbf{k+\mathbf{q}}})
(\psi^{\dag}_{\mathbf{k^{\prime}+\mathbf{q}}}\tau_3\psi_{\mathbf{k}^{\prime}})
\,,
\end{equation}
where $g<0$ is a coupling constant.

The correlator $K^{ij}=-\langle\mathrm{ Tr}[\mathrm{T}\left(\sigma^i_1(\tau)\sigma^j_2(0)\right)]\rangle$ of spin densities in Eq.(\ref{U}) depends on the impurity coordinate difference $\mathbf{r}\equiv(\mathbf{r}_1-\mathbf{r}_2)$. Therefore, it may be Fourier transformed with respect to $\mathbf{r}$. The so transformed correlator, which is integrated over the imaginary time $\tau$ in Eq.(\ref{U}), will be denoted as $K^{ij}_{\mathbf{q}}$, where $\mathbf{q}$ is the wave-vector. Within a perturbation theory \cite{AGD} it may be further expanded into the series with respect to electron-electron interaction Eq.(\ref{V}). Within the BCS self-consistent approximation the perturbation terms contribute to the self-energy of the one-particle Matsubara Green's functions $G_{\mathbf{k}}(\omega_n)$ and result in the superconducting order parameter $\Delta$, so that $G_{\mathbf{k}}(\omega_n)=(i\omega_n-\mathcal{H}_{0\mathbf{k}}-\tau_1\Delta)^{-1}$. In a simplest  approximation one may take $V$ into account only in the self-energy of $G$ and calculate $K^{ij}_{\mathbf{q}}$ with such BCS Green's functions. The corresponding correlator, which will be denoted as $K^{ij}_{\mathrm{RKKY}}$ can be written in the form $K^{ij}_{\mathrm{RKKY}}=4\Pi_{00}^{ij}(\mathbf{q})$, where
\begin{equation}\label{Pi}
\Pi_{lm}^{ij}(\mathbf{q})=\frac{T}{4}
\sum_{\omega_n}\mathrm{Tr}[G_{\mathbf{k}}(\omega_n)\sigma^i\tau_lG_{\mathbf{k+q}}(\omega_n)\sigma^j\tau_m] \,.
\end{equation}
The superscripts $i$ and $j$ take the values $x,y,z$ and 0, where the latter denotes the  2$\times$2 unit matrix $\sigma^0$. The Green functions in Eq.(\ref{Pi}) consist of two parts which are associated with one of the helical bands. It will be assumed below that the SOC is much larger than $\Delta$. In this case the main contribution in Eq.(\ref{Pi}) is given by the terms where both Green functions enter with equal helicities.  The details of the calculation are presented in Appendix A.

By substituting into Eq.(\ref{U}) $K^{ij}_{\mathrm{RKKY}}$ results in the conventional RKKY interaction, which has been previously calculated for s-wave superconductors, with and without SOC \cite{Anderson,Zyuzin,Qin}. It takes account of quasiparticles as mediators of the magnetic impurity interactions. At the same time, perturbations of the condensate by these impurities are ignored in this approximation. They can be taken into account beyond the simple BCS approach, by summing up in the perturbation expansion the so called ladder Feynman diagrams,  which contain a two-particle Cooper pairing channel. \cite{Nambu} These diagrams represent multiple scattering amplitude of two particles at the Fermi energy. These particles are in a singlet spin state and have opposite momenta, while the conserved wave-vector $\mathbf{q}$ describes the dependence of the two-particle wave-function on the center of gravity. The so called "Cooperon" $C_{\mathbf{q}}$, which is the sum of the ladder diagrams, can be expressed through the function $\Pi_{22}^{00}(\mathbf{q})$ in the form (see Appendix A)
\begin{equation}\label{C}
C_{\mathbf{q}}=-\frac{g}{1-g\Pi_{22}^{00}(\mathbf{q})}\,.
\end{equation}
Note, that at $\mathbf{q}=0$ the denominator of Eq.(\ref{C}) turns to zero, because of the selfconsistency equation $g\Pi^{00}_{22}(0)=1$. The divergence of $C_{\mathbf{q}}$  at  $q\rightarrow 0$ is due to the invariance of the system with respect to a static uniform shift of the order parameter phase. As it will be shown, this singularity results in a long-range interaction of impurity spins. It is important that the static situation is considered. Therefore, the long-range Coulomb interaction does not destroy the $q=0$ singularity, in contrast to the dynamical Goldstone mode. Formally, it follows from the fact that at zero frequency the interaction of the Cooperon with the Coulomb field turns to zero \cite{Nambu}.

The Cooperon represents a correlation of two electrons having a zero total spin. At the same time, in a spin-orbit coupled system one may expect that SOC can provide a link between spin singlet and spin triplet states which, in turn, can interact with magnetic impurities. Within a perturbation theory formalism such a link is represented by the vertex function $\Gamma^i_{\mathbf{q}}$, which is given by
\begin{equation}\label{Gamma}
\Gamma^i_{\mathbf{q}}=2\Pi_{02}^{i0}(\mathbf{q})\,.
\end{equation}
Note that the nondiagonal "$i0$" superscript in $\Pi_{02}^{i0}(\mathbf{q})$ signals that $\Gamma^i_{\mathbf{q}}$ couples the spin projection $i$ to a singlet Cooper pair.  In terms of  the vertex function and Cooperon the correlator $K^{ij}_{\mathbf{q}}$ can be written  in the form
\begin{equation}\label{K}
K^{ij}_{\mathbf{q}}=K^{ij}_{\mathrm{RKKY}}+\Gamma^i_{\mathbf{q}}C_{\mathbf{q}}\Gamma^j_{-\mathbf{q}}\,.
\end{equation}
Two vertices $\Gamma^i_{\mathbf{q}}$ and $\Gamma^j_{-\mathbf{q}}$ provide a coupling of the Cooperon to spins placed at points $\mathbf{r}_1$ and $\mathbf{r}_2$.

By expanding $\Pi_{22}^{00}(\mathbf{q})$ in Eq.(\ref{C}) up to $q^2$  we obtain
\begin{equation}\label{C2}
C_{\mathbf{q}}=-\frac{a}{q^2}\frac{16}{v_F^2}(N_{F+}+N_{F-})^{-1}\,,
\end{equation}
where
\begin{equation}\label{a}
\frac{1}{a}=\pi T\sum_{\omega_n}\frac{1}{(\omega_n^2+\Delta^2)^{3/2}}\,.
\end{equation}
The vertex $\Gamma$ can be expressed from Eqs.(\ref{Gamma},\ref{Pi}) as
\begin{equation}\label{Gamma2}
\Gamma^i_{\mathbf{q}}=-\frac{i}{4a}\tilde{q}^i\Delta v_F  (N_{F+}-N_{F-})\,,
\end{equation}
where $\tilde{q}^i=\epsilon^{zij}q^j$. By substituting Eqs.(\ref{C2}) and (\ref{Gamma2}) in Eq.(\ref{K}) we finally obtain
\begin{equation}\label{K2}
K^{ij}_{\mathbf{q}}-K^{ij}_{\mathrm{RKKY}}=-\frac{\tilde{q}^i\tilde{q}^j}{q^2}\frac{\Delta^2 (N_{F+}-N_{F-})^2}{a(N_{F+}+N_{F-})}\,.
\end{equation}
It is seen from this equation that two helical bands tend to compensate each other, while the above expression reaches its maximum in the case when there is only a single helical band on the Fermi surface, for example, in the case of Dirac electrons on the surface of a topological insulator.

\section{Long-range interaction between magnetic impurities}

By Fourier transforming Eq.(\ref{K2}) to the coordinate representation and substituting  into Eq.(\ref{U}) we obtain the condensate mediated interaction between magnetic impurities, which are  placed at the sites $\mathbf{r}_1$ and $\mathbf{r}_2$, in the form
\begin{equation}\label{U2}
U_{12}=\frac{\beta}{r^4}\left(2(\mathbf{Z}_1\tilde{\mathbf{r}})(\mathbf{Z}_2\tilde{\mathbf{r}}) - r^2\mathbf{Z}_{1\parallel}\mathbf{Z}_{2\parallel}\right)\,,
\end{equation}
where $\beta=(\Delta^2/2\pi a) (N_{F+}-N_{F-})^2(N_{F+}+N_{F-})^{-1}$, $\tilde{r}^i=\epsilon^{zij}r^j$, with $r^j=r^j_1-r^j_2$ and $\mathbf{Z}_{1(2)\parallel}=(Z_{1(2)}^x,Z_{1(2)}^y)$. Eq.(\ref{U2}) is valid at  $r\gg v_F/\Delta$. For a chain of magnetic atoms, whose spins are perpendicular to the chain direction, the interaction is antiferromagnetic, while it is ferromagnetic for spins which are parallel to the chain. For a ferromagnetically ordered two-dimensional array of spins the long-range interaction results in the mean field $B$ which depends on the shape of the system. For example, in the center of  a rectangle whose sides are $L_x$ and $L_y$   this field may be expressed from Eq.(\ref{U2}) by summing $U$ over all thermodynamically averaged spins $\langle S_{2x}\rangle$, with the spin $S_1$ fixed in the rectangle center. Since mostly distant spins  contribute to this sum, one may replace the summation by integration with the spin density $n_i$. By writing the result  in the form $\sum_m U_{1m}=S_{1x} B_x$ the mean field $B_x$ is obtained as
\begin{equation}\label{B}
B_x=2n_i \beta J^2\langle S_{2x} \rangle \left(\arctan\frac{L_y}{L_x} - \arctan\frac{L_x}{L_y}\right)\,,
\end{equation}
where the exchange interaction is taken as $J_{ij}=\delta_{ij}J$. It is evident from this equation that the ferromagnetic ordering is possible only at $L_x> L_y$. The dependence of the mean field on the shape of the sample is a consequence of the formal logarithmic divergence of the mean field, which is caused by a slow decrease of the impurity spins interaction at large distances. Such a divergence does not take place in the case of one dimensional spin chains. However, in two dimensional systems it can strongly influence the spin order.

$U_{12}$ in Eq.(\ref{U2}) does not contain  the Dzyaloshinskii-Moryia interaction. In fact, this interaction appears only in higher orders with respect to $1/rk_F $. Such small corrections were neglected above. On the other hand, the Dzyaloshinskii-Moryia term might appear also in the case when, due to SOC, the attractive electron-electron interaction $V$ in Eq.(\ref{V}) contains a spin-dependent term. This possibility has not been investigated here.

\section{Bilayer system}

From the  practical point of view it is important to consider a system where superconductivity is induced in a spin-orbit coupled normal metal due to a close proximity to a superconductor. In this case a popular approach is to insert into  Hamiltonian of the normal system a term which looks as the superconducting order parameter. In some cases such an approach is justified. It is definitely can not be used in the studied here problem, because the selfconsistency condition for the order parameter is determined by an electron-electron attractive interaction inside the superconductor, not the normal metal. Therefore, if a pair of magnetic impurities interacts through the condensate, the superconductor must explicitly be considered as a mediator for this interaction. A strategy for the solution of this problem may be to calculate the  spin density $\langle\sigma^i_{21}\rangle$ which is induced at the point $\mathbf{r}_2$ by a magnetic impurity paced in the point $\mathbf{r}_1$, and vice versa. Then, the interaction energy of these spins can be expressed as
\begin{equation}\label{U3}
U_{12}= J_{ij}(S^i_{1}\langle\sigma^j_{12}\rangle +S^i_{2}\langle\sigma^j_{21}\rangle)\,.
\end{equation}
The induced spin densities must be calculated by taking into account a perturbation of the condensate by magnetic impurities. It is important that in contrast to  Eq.(\ref{U}), where the interaction energy is determined by a two-particle Green function, in Eq.(\ref{U3}) the spin densities are calculated through the one-particle Matsubara Green's functions $G(\omega_n,\mathbf{r},\mathbf{r}^{\prime})$.

Let us consider a bilayer system consisting of a superconducting layer and an adjacent 2D normal metal. The latter is  formed by Dirac electrons on the surface of a 3D TI. Their Hamiltonian is represented by the second term  in Eq.(\ref{H0}). A pair of magnetic impurities on the surface of TI adds the Zeeman term $\bm{\sigma}\mathbf{Z}(\mathbf{r})$, where $\mathbf{Z}=\mathbf{Z}_1\delta(\mathbf{r}-\mathbf{r}_1)+\mathbf{Z}_2\delta(\mathbf{r}-\mathbf{r}_2)$. For studying such an inhomogeneous system we will use a theory based on semiclassical Green's functions. These functions are defined separately in both layers of the considered bilayer system and satisfy respective semiclassical equations \cite{Eilenberger,Larkin semiclass}. In addition, there is a boundary condition at the interface between layers.  The semiclassical function is defined as
\begin{equation}\label{gnu}
g_{\hat{\mathbf{k}}}(\mathbf{r},\omega)=\frac{i}{\pi}\int d\xi \tau_3G_{\mathbf{k}}(\mathbf{r},\omega_n)\,,
\end{equation}
where $\xi=E_{\mathbf{k}}-\mu$, $\mathbf{r}=(\mathbf{r}+\mathbf{r}^{\prime})/2$, $\mathbf{k}$ is the wave vector associated with the Fourier transform of $G(\omega_n,\mathbf{r},\mathbf{r}^{\prime})$ with respect to $\mathbf{r}-\mathbf{r}^{\prime}$ and $\hat{\mathbf{k}}=\mathbf{k}/k$. The electron energy $E_{\mathbf{k}}$ is given either by a parabolic $k$-dependence (in the superconducting layer), or by a linear function $h_{\mathbf{k}}$ (in TI).  The  equation for $g_{\hat{\mathbf{k}}}(\mathbf{r},\omega)$ is obtained by expanding the Dyson equation with respect to  Fermi wavelengths, which are small in comparison with other characteristic lengths. The corresponding procedure is well described in literature \cite{Larkin semiclass,Rammer, Kopnin}. It will be assumed below that the elastic scattering time  on impurities $\tau \ll \Delta^{-1}$ and the corresponding mean free path is much shorter than the lengths of spatial variations of the semiclassical Green functions.
In this case these functions are almost isotropic with respect to $\hat{\mathbf{k}}$ and it is possible to obtain closed, so called, Usadel \cite{Usadel,Larkin semiclass,Rammer, Kopnin} equations for their isotropic parts $g_{S(N)}(\mathbf{r},\omega)$, where subscripts N and S denote the normal and superconducting layers, respectively. For the S-layer such an  equation can be written in the standard form
\begin{equation}\label{Usadel1}
D_S\bm{\nabla}g_S\bm{\nabla}g_S - [\omega\tau_3+\hat{\Delta},g_S]=0\,,
\end{equation}
where $\hat{\Delta}=\texttt{Re}\Delta(\mathbf{r})\tau_2-\texttt{Im}\Delta(\mathbf{r})\tau_1$. The equation for TI can be obtained by a projection of $g_{N\hat{\mathbf{k}}}$  onto the upper helix band, if the chemical potential $\mu>0$ and $\mu\gg 1/\tau$ \cite{ZyuzinTI,Bobkova,Linder}. Accordingly, $g_{N\hat{\mathbf{k}}}$ takes the form $g_{N\hat{\mathbf{k}}}=g_{N\hat{\mathbf{k}}0}(1+\bm{\sigma}\mathbf{n})/2$, where $\mathbf{n}=\mathbf{h}_{\mathbf{k}}/h_{\mathbf{k}}$ and $g_{N\hat{\mathbf{k}}0}$ is a spin independent function. Its angular average  $g_{N}$ satisfies the equation
\begin{equation}\label{Usadel2}
D_N\tilde{\bm{\nabla}}g_{N}\tilde{\bm{\nabla}}g_{N}-[\omega\tau_3+T_{NS}g_S,g_{N}]=0 \,,
\end{equation}
where $\tilde{\bm{\nabla}}*=\bm{\nabla}*+i[\mathbf{A(\mathbf{r})}\tau_3,*]$ and $A^i(\mathbf{r})=\epsilon^{ijz}Z^j(\mathbf{r})/\alpha$. $D_S$ and $D_N$ are electron diffusion coefficients in the superconductor and normal layers, respectively. The last term in Eq.(\ref{Usadel2}) originates from the self-energy associated with a tunnel coupling of 2D electrons to the superconducting layer, where $T_{NS}$ is a corresponding tunneling parameter \cite{bilayer}. From the superconductor's side the coupling to the normal layer is provided by the boundary condition (BC) \cite{Kupriyanov}
\begin{equation}\label{BC}
D_Sg_S\nabla_zg_S=-\gamma_{SN}[g_S,g_{N}]\,,
\end{equation}
where the $z$-axis is directed from the $N$-layer to the $S$-layer, $g_S$ is taken at $z=0$ and $\gamma_{NS}$ can be expressed in terms of the interface resistance.

The Usadel equation for $g_S$ may be further simplified \cite{Zaitsev} by assuming that Green functions vary slowly  across a thin film, whose thickness $d_S$ is much less than the superconductor's coherence length $\sqrt{D_S/|\Delta|}$. By integrating Eq.(\ref{Usadel1}) over $z$ and taking into account BC Eq.(\ref{BC}) we obtain the following equation for  $g_S(\mathbf{r})=(1/d_S)\int dz g_S(\mathbf{r},z)$:
\begin{equation}\label{UsadelS}
D_S\bm{\nabla}g_S\bm{\nabla}g_S -[\omega\tau_3+\hat{\Delta}+T_{SN}g_N,g_S]=0\,,
\end{equation}
where $T_{SN}=D_S\gamma_{SN}/d_S$. The parameters $T_{NS}$ and $T_{SN}$ are related to each other through the equation $N_{F_N}T_{NS}=d_SN_{F_S}T_{SN}$, where $N_{F_N}$ and $N_{F_S}$ are, respectively, 2D and 3D state densities at the Fermi level in the normal metal and superconductor (in the normal state). This equation guarantees the conservation of the charge current through the NS-interface. It should be noted that the above relation between $T_{NS}$ and $T_{SN}$ means that  $T_{SN} \ll T_{NS}$, because $k_{FS}d_S\gg 1$.

As was shown in \cite{Malsh island,Pershoguba}, in spin-orbit coupled superconductors a Zeeman field, which is localized within a small island, induces in its vicinity a spontaneous supercurrent. A single magnetic impurity may produce a similar effect. This supercurrent, in turn, induces a spin density due to the magnetoelectric effect \cite{Edelstein,EdelsteinPRL} which takes place in spin-orbit coupled superconductors. The interaction of impurity spins may be calculated by substituting this spin density into Eq.(\ref{U3}). There are two different mechanisms which contribute in the formation of such a spin-density  response. One of them involves the supercurrent, which is directly produced by proximity induced Cooper pairs. This effect is controlled by a proximity  induced gap $\Delta_N$. The latter coincides with $T_{NS}$ when  $T_{NS}\ll \Delta$. As a result, the spin interaction given by Eq.(\ref{U3}) decreases exponentially at distances larger than the corresponding correlation length $\xi_N=$min$[\sqrt{D_N/\Delta_N},\sqrt{D_N/2\pi k_BT}]$. The second mechanism involves several steps. An important step is a change of the superconducting order-parameter in the S-layer, due to a perturbation of the pairing function by the Zeeman field. This perturbation migrates from the normal layer through the interface barrier. The correction to $\Delta$  may be expressed in a form of a phase shift. The latter gives rise to a supercurrent in  N and S layers and to a spin polarization, which  results in the interaction of magnetic impurities. These calculations in detail  are presented in Appendix B. The main result is that at large distances $r \gg \xi_N$  the interaction of spins is expressed by Eq.(\ref{U2}), where at $T\ll \Delta_N,\Delta$ the coefficient $\beta$ is given by
\begin{equation}\label{beta}
\beta=\frac{\mu\tau T_{NS}T_{SN} D_N}{2\pi\alpha^2 \Delta D_S}\,.
\end{equation}
This $\beta$ is much less than in the considered above case of a clean 2D superconductor. It is strongly reduced by the factor $d_Sk_{FS} \gg 1$. This parameter enters into $T_{SN}$. The origin of such a reduction is quite clear, because the influence of the SOC in the  2D gas on the condensate in the 3D superconductor film decreases  at the larger film thickness $d_S$.

\section{Conclusion}

In conclusion, in spin-orbit coupled 2D superconductors, or 2D normal Dirac metals which have a contact with thin superconducting films, the interaction of magnetic impurity spins decreases as $r^{-2}$ at distances larger than the coherence length. In fact, it has the form of the 2D dipole-dipole interaction. In contrast to the  exponentially decreasing RKKY interaction, it is mediated by the Cooper pair condensate, rather than by one-particle excitations. In the case of proximity induced superconductivity in a 2D normal metal this long-range effect is suppressed, if a 3D superconductor, which serves as a source of Cooper pair correlations, is massive.

\emph{Acknowledgements} - The work was partly supported by the Russian Academy of Sciences program "Actual
 problems of low-temperature physics."


\appendix

\section{Interaction of impurity spins in a 2D superconductor}

The BCS Green function of a system which is represented by the Hamiltonian $\mathcal{H}_{0\mathbf{k}}+\tau_1\Delta$, where the first term is given by  Eq.(2), can be written as
\begin{eqnarray}\label{G}
G_{\mathbf{k}}(\omega_n) = && - \frac{{(i\omega_n  + {\tau _1}\Delta  + {\xi ^ + }{\tau _3}})}{{{\omega^2_n} + {{({E^ + })}^2}}}\frac{({1 + {\bf{n}_{\mathbf{k}}}\bm{\sigma} })}{2} -  \nonumber \\
&&\frac{({i\omega_n  + {\tau _1}\Delta  + {\xi ^ - }{\tau _3}})}{{{\omega ^2_n} + {{({E^ - })}^2}}}\frac{({1 - {\bf{n}_{\mathbf{k}}}\bm{\sigma} })}{2}\,,
\end{eqnarray}
where $E^{\pm}=\sqrt{\Delta^2+\xi ^ {\pm 2}}$,  $\xi ^ {\pm}=\xi \pm h_{\mathbf{k}}$, $\xi=\epsilon_{\mathbf{k}}-\mu$ and $\mathbf{n}_{\mathbf{k}}=\mathbf{h}_{\mathbf{k}}/h_{\mathbf{k}}$. These functions must be substituted into the correlator given by Eq.(4). This correlator is the main building block of the Cooperon, which is represented by a sum of ladder diagrams, where the electron-electron interaction $V$, given by  Eq.(3), serves as a perturbation. The dependence of this interaction on Nambu and spin variables can be represented in the form
\begin{equation}\label{V2}
V= \frac{1}{2} \sum_{\mathbf{k,k^{\prime},q}}V_{\alpha\beta,\gamma\delta}\psi^{\alpha\dag}_{\mathbf{k}+\mathbf{q}}\psi^{\gamma\dag}_{\mathbf{k^{\prime}}}\psi^{\beta}_{\mathbf{k}^{\prime}+\mathbf{q}}
\psi^{\delta}_{\mathbf{k}}\,,
\end{equation}
where $V_{\alpha\beta,\gamma\delta}=(g/2)(\tau_3\otimes\sigma^0)_{\alpha\beta}(\tau_3\otimes\sigma^0)_{\gamma\delta}$. The Greek subscripts denote combined Nambu-spin variables. It is convenient to use, instead, vector indices by transforming the matrix $V$ to
\begin{equation}\label{Vij}
V^{ij}_{ab}=\frac{1}{4}V_{\alpha\beta,\gamma\delta}(\tau_a\otimes\sigma^i)_{\gamma\alpha}(\tau_b\otimes\sigma^j)_{\beta\delta}\,,
\end{equation}
where $i,j=0,x,y,z$ and $a,b=0,1,2,3$.

In this representation, the ladder series for the two-particle scattering matrix  $C^{ij}_{pq}(\mathbf{q})$ can be written in the form of the following  equation \cite{AGD}:
\begin{equation}\label{Cij}
C^{ij}_{ab}=V^{ij}_{ab}+V^{ik}_{ac}\Pi^{kl}_{cd}(\mathbf{q})C^{lj}_{db}\,,
\end{equation}
where the argument $\mathbf{q}$ is omitted in $\mathcal{C}$ for brevity. Superscripts in this equation relate to spin variables in the vector representation, while the subscripts are associated with the Nambu space. We assume that SOC is strong, so that the spin-orbit field $h_{\mathbf{k}_F} \gg \Delta$. In this case, the poles of the first and second terms  in Eq.(\ref{G}), which correspond to opposite chiralities, are considerably shifted with respect to each other. Therefore, the main contribution in pair products of Green's function in $\Pi$ (Eq.(\ref{Pi})) is given by the functions with equal chiralities. Hence, the products of Green's functions with opposite chiralities will be neglected. Further, we consider the situation when a distance between magnetic impurities is much larger than the Fermi wavelength. Therefore,  the Green function $G_{\mathbf{k}+\mathbf{q}}$, which enters in $\Pi$, may be expanded with respect to $q \ll k_F$. Only the leading terms of this expansion will be taken into account. As can be seen from Eq.(\ref{Pi}) and Eq.(\ref{G}), the spin dependent part of  $\Pi^{ij}_{pq}$ is given by $\mathrm{Tr}[(1\pm \mathbf{n}_{\mathbf{k+q}}\bm{\sigma})\sigma^i (1\pm \mathbf{n}_{\mathbf{k}}\bm{\sigma})\sigma^j]$. This trace, averaged over directions of the unit vector $\mathbf{n}_{\mathbf{k}}$, is nonzero only at $i=j$. At the same time, the nonaveraged trace contains nondiagonal terms of the form $\pm 4n^i_{\mathbf{k}}$ at $j=0$, or $\pm 4n^j_{\mathbf{k}}$ at $i=0$ (small terms of the order of $q/k_F$  are ignored). By tracing out the spin-dependent part in the diagonal elements, the remaining calculation of the sum over $\mathbf{k}$ in Eq.(\ref{Pi}) is reduced to the conventional  analysis \cite{Nambu}.  It follows then that the singular at $q\rightarrow 0$ Cooper pairing channel corresponds to the spin-singlet Cooperon $C^{00}_{22}$ and, hence,  involves the correlator $\Pi^{00}_{22}$, which is given by
\begin{eqnarray}\label{Pi2}
&&\Pi^{00}_{22}(\mathbf{q})=-\pi T\frac{N_{F+}+N_{F-}}{2}\left(\sum_{\omega_n<\omega_c}\frac{1}{\sqrt{\omega_n^2+\Delta^2}}\right. +\nonumber \\
&& \left. \frac{q^2v_F^2}{8}\sum_{\omega_n}\frac{1}{\sqrt{\omega_n^2+\Delta^2}^3}\right)
\end{eqnarray}
Let us first keep in Eq.(\ref{Cij}) only this leading term and take into account that, according to Eqs.(\ref{V2}) and (\ref{Vij}), $V^{ij}_{00}=V^{ij}_{33}=g\delta^{ij}$ and $V^{ij}_{11}=V^{ij}_{22}=-g\delta^{ij}$.
As a result, the equation for $\mathcal{C}^{00}_{22}$ in Eq.(\ref{Cij}) is decoupled from equations for other components of $\mathcal{C}$ and leads to Eq.(\ref{C}), where $\mathcal{\mathcal{C}}_{\mathbf{q}}\equiv\mathcal{C}^{00}_{22}(\mathbf{q})$.

Let us now analyze the effect of the neglected nondiagonal terms, which give rise to $\Pi^{0i}_{20}(\mathbf{q})$ and $\Pi^{i0}_{02}(\mathbf{q})$. These nondiagonal in spin and Nambu spaces correlators appear due to SOC. By using Eq.(\ref{Pi}) and Eq.(\ref{G}), in the leading with respect to $q$ approximation $\Pi^{0i}_{20}(\mathbf{q})$ can be written as
\begin{eqnarray}\label{Pi20}
&&\Pi^{0i}_{20}(\mathbf{q})=2iT\sum_{\omega_n,\mathbf{k}}n^i_{\mathbf{k}}\Delta \left(\frac{\xi^+_{\mathbf{k+q}}-\xi^+_{\mathbf{k}}}{(\omega^2_n + E^ {+2}_{\mathbf{k}})^2}- \right. \nonumber \\
&&\left. \frac{\xi^-_{\mathbf{k+q}}-\xi^-_{\mathbf{k}}}{(\omega^2_n + E^ {-2}_{\mathbf{k}})^2}\right)
\end{eqnarray}
By representing $(\xi^{\pm}_{\mathbf{k+q}}-\xi^{\pm}_{\mathbf{k}})$ as $\mathbf{v_Fq}$  and taking the sum over $\mathbf{k}$ we arrive at
\begin{equation}\label{Pi202}
\Pi^{0i}_{20}(\mathbf{q})=\frac{i}{8a}\epsilon^{zij}q^j\Delta v_F  (N_{F+}-N_{F-})\,,
\end{equation}
where $a$ is given by Eq.(\ref{a}). $\Pi^{0i}_{20}(\mathbf{q})$ is small in comparison with $\Pi^{00}_{22}(\mathbf{q})$. It  can be taken into account in Eq.(\ref{Cij}) as a small perturbation. It is easy to see that it results in $\mathcal{C}^{00}_{22}$, which is given by Eq.(\ref{C}) with  $\Pi^{00}_{22}(\mathbf{q})$ substituted for $\Pi^{00}_{22}(\mathbf{q})+\Pi^{0i}_{20}(\mathbf{q})V_{00}^{ii}\Pi^{i0}_{02}(\mathbf{q})$. It is seen that such a correction to $\Pi^{00}_{22}(\mathbf{q})$ is small as $gv_F^2q^2/\Delta^2 $. Since the coupling parameter $g\ll 1$, this correction  is small in comparison with the term of the order of  $v_F^2q^2/\Delta^2$, which originates from a direct expansion of $\Pi^{00}_{22}(\mathbf{q})$ in powers of $q$ and appears in the denominator of Eq.(\ref{C2}) (it is given by the second term in Eq.(\ref{Pi2}). Therefore, one may neglect a contribution of the nondiagonal elements of $\Pi$.

By taking into account only Cooper pair correlations in $K^{ij}_{\mathbf{q}}$ one can express the latter in the form
\begin{equation}\label{Kij}
K^{ij}_{\mathbf{q}}-K^{ij}_{\mathrm{RKKY}}=4\Pi^{i0}_{02}(\mathbf{q})C^{00}_{22}(\mathbf{q})\Pi^{0j}_{20}(\mathbf{q})\,,
\end{equation}
or, equivalently, in the form of Eq.(\ref{K}).

\section{Interaction of impurity spins in a bilayer system}

Let us focus on Eqs.(17) and (19) for semiclassical Green's functions in a bilayer system which consists of a superconducting layer and a normal 2D Dirac electron gas. The spatially dependent Zeeman field, which enters into the gauge-invariant derivative in Eq.(17), will be treated as a perturbation and only first-order terms will be taken into account. Therefore, it is convenient to transform these equations into the Fourier representation. Within this perturbation approach the semiclassical Green functions and the order parameter can be represented in the form
\begin{eqnarray}\label{deltaG}
g_S=&&g_{S0}+\delta g_{S}\,\,\,,\,\,\,g_N=g_{N0}+\delta g_{N}\nonumber \\
&&\text{and}\,\,\,\hat{\Delta}=\hat{\Delta}_0+\delta\hat{\Delta}\,.
\end{eqnarray}
The unperturbed functions $g_{S0}$, $g_{N0}$ and the order parameter $\hat{\Delta}_{0}$ are uniform along layers. They are determined by Eqs.(19) and (17) in the absence of magnetic impurities. Since the tunneling parameter $T_{SN}$ is small, one may ignore a weak influence of the 2D normal gas onto the Green's function and the order parameter of the superconducting film. Therefore, we take them the same as in a bulk superconductor, namely
\begin{equation}\label{g0Delta0}
g_{S0}=\frac{\omega_n\tau_3+\Delta_0\tau_2}{\sqrt{\omega_n^2+\Delta_0^2}}\,\,\,\text{and}\,\,\,\hat{\Delta}_0=\Delta_0\tau_2\,.
\end{equation}
At the same time, the superconductor proximity effect leads to important changes in  the Green's function of the normal gas. The corresponding uniform solution $g_{N0}$ can by obtained from the second term of Eq.(17), which must be zero. Hence, $g_{N0}$ is given by
\begin{equation}\label{gN0}
g_{N0}=\frac{\omega_n\tau_3+T_{NS}g_{S0}}{\sqrt{(\omega_n\tau_3+T_{NS}g_{S0})^2}}\,.
\end{equation}
This function satisfies the normalization condition $g_{N0}^2=1$. At $\omega_n \ll \Delta$ it takes the form of Eq.(\ref{g0Delta0}) with $\Delta_0$ substituted for $\Delta_N\equiv T_{NS}$. Hence, $g_{N0}$  looks as the Green's function of a superconductor, where the role of the gap is played by $\Delta_N$.

The corrections due to the Zeeman field are obtained from linearized Eqs.(19) and (17). They are given by
\begin{eqnarray}\label{deltaSN}
\delta g_{S}&=&\frac{1}{D_Sq^2+2\Omega_S^2}\left(T_{SN}g_{S0}[g_{S0},\delta g_{N}]-g_{S0}[\delta\hat{\Delta},g_{S0}] \nonumber \right)\\
\delta g_{N}&=&-\frac{1}{D_Nq^2+2\Omega_N^2}\left(T_{NS}g_{N0}[\delta g_{S},g_{N0}]+\nonumber \right.\\
&&\left. D_N(\mathbf{qA})[\tau_3,g_{N0}]\right) \,,
\end{eqnarray}
where $\Omega_S=\omega_n\tau_3+\Delta_0\tau_2$ and $\Omega_N=\omega_n\tau_3+T_{NS}g_{S0}$. It is seen from Eq.(\ref{gN0}) that the term in Eq.(\ref{deltaSN}), which is associated with the Zeeman field (the second term in the second equation), is proportional to the Pauli matrix $\tau_1$. Then, it is easy to see that $\delta g_{N},\delta g_{S}$ and $\delta\hat{\Delta}$ are also proportional to $\tau_1$. Accordingly, we denote $\delta \hat{g}_{N}=\tau_1\delta g_{N},\delta \hat{g}_{S}=\tau_1\delta g_{S}$, and $\delta\hat{\Delta}= \tau_1\delta\Delta$. Therefore, let us project Eq.(\ref{deltaSN}) onto $\tau_1$. By resolving these equations we obtain
\begin{eqnarray}\label{deltaSN2}
\delta g_{S}&=&\frac{2}{\mathcal{D}}\left(b_N \delta\Delta +2iT_{SN}D_N(\mathbf{qA})g^{(2)}_{N0}\right) \nonumber \\
\delta g_{N}&=&\frac{2}{\mathcal{D}}\left(2T_{NS} \delta\Delta+ib_SD_N(\mathbf{qA})g^{(2)}_{N0}\right)\,,
\end{eqnarray}
where $b_{N(S)}=D_{N(S)}q^2+2\Omega_{N(S)}^2$, $g^{(2)}_{N0}=\mathrm{Tr}[\tau_2 g_{N0}]$, and $\mathcal{D}=b_Nb_S-4T_{SN}T_{NS}$.

The selfconsistency reads
\begin{equation}\label{deltaS}
 \delta\Delta=\frac{gN_{F_S}T}{2}\sum_{\omega_n}\delta g_{S}\,.
\end{equation}
The correction  $\delta\Delta$ to the gap can be obtained by substituting in this equation $\delta g_{S}$ from Eq.(\ref{deltaSN2}). We will keep only leading terms in $T_{SN}$. By taking into account the unperturbed selfconsistency equation $1=gN_{F_S}T\sum_{\omega_n}(1/2|\Omega_s|)$ this correction may be expressed from  Eq.(\ref{deltaS}) in the form
\begin{equation}\label{deltaS2}
 \delta\Delta=4T_{SN}\frac{i\mathbf{qA}}{q^2}\frac{D_N}{D_S}\left(\sum_{\omega_n}\frac{g^{(2)}_{N0}}{b_Nb_S}\big{/}\sum_{\omega_n}\frac{1}{|\Omega_s|b_S}\right)\,.
\end{equation}
This expression should be substituted into Eq.(\ref{deltaSN2}). Further, $\delta g_{N}$ can be used for the calculation of the spin densities $\langle\sigma^j_{12}\rangle$ and $\langle\sigma^j_{21}\rangle$ in Eq.(14). For example, the former is given by
\begin{eqnarray}\label{sigma12}
&&\langle\sigma^j_{12}\rangle =\frac{T}{2}\sum_{\omega_n,\mathbf{k}}\mathrm{Tr}[\sigma^j G_{\mathbf{k}}(\mathbf{r_1},\omega_n)]=\nonumber \\&&-i\frac{\pi}{2}TN_{F_N}\sum_{\omega_n}\int \frac{d\phi}{2\pi}\mathrm{Tr}\left[\tau_3\sigma^j\frac{1+\mathbf{n}\bm{\sigma}}{2}\delta g_{N\hat{\mathbf{k}}}(\mathbf{r}_1)\right]\,,
\end{eqnarray}
where the angle $\phi$ specifies a direction of the unit vector $\mathbf{\hat{k}}$. Due to the trace over spin variables, the integrand is proportional to $n^j$, which is antisymmetric with respect to $\mathbf{\hat{k}}$. At the same time, the antisymmetric in $\mathbf{\hat{k}}$ function $\delta g_{N\hat{\mathbf{k}}}$ may be expressed in terms of the angular-averaged function $\delta g_{N}$, according to $\delta g_{N\hat{\mathbf{k}}}=-2\tau \alpha \mathbf{\hat{k}}g_{N0}\tilde{\bm{\nabla}}\delta g_{N}(\mathbf{r}_1)$ [24-26]. Within the linear in $\mathbf{Z}$ approximation,  in the latter expression $\tilde{\bm{\nabla}}$ may be substituted for $\bm{\nabla}$, while in $\delta g_{N}(\mathbf{r}_1)$ only the Zeeman field $\mathbf{Z}_2\delta(\mathbf{r}-\mathbf{r}_2)$ must be taken into account. By substituting $\delta g_{N\hat{\mathbf{k}}}(\mathbf{r}_1)$ into Eq.(\ref{sigma12}) and calculating the trace over spin and Nambu variables we obtain
\begin{equation}\label{sigma122}
\langle\sigma^j_{12}\rangle=-i\pi TN_{F_N}\tau\sum_{\omega_n,\mathbf{q}}\tilde{q}^jg_{N0}^{(2)}\delta g_{N}e^{i\mathbf{q}(\mathbf{r}_{1}-\mathbf{r}_{2})}\,.
\end{equation}

The function $\delta g_{N}$ in Eq.(\ref{sigma122}) contains two parts, as can be seen from Eq.(\ref{deltaSN2}). One of them is proportional to $ \delta\Delta$. It stems from the condensate, which is perturbed by magnetic impurities. The second term represents a direct effect of  the Zeeman field, where the sole role of the superconductor is  to produce the proximity gap $\Delta_N$ in the normal film. Indeed, by neglecting the small term in $\mathcal{D}$, which is proportional to $T_{SN}$, at $T_{NS}\ll \Delta$, one obtains for $\delta g_{N}$ (given by the second term in Eq.(\ref{deltaSN2})) the same expression as in the case of a superconductor whose energy gap is $\Delta_N=T_{NS}$. It follows straight from  Eq.(\ref{gN0}) which at $\omega_n$ and $T_{NS} \ll \Delta$ takes the form of the semiclassical Green function of a superconductor. Therefore, this part of $\delta g_{N}$ contributes to the "smooth" RKKY interaction [14] (the oscillating with $k_Fr$ term can not be treated within the semiclassical theory). The dependence on a distance between magnetic impurities is seen from the pole of $\delta g_{N} \sim 1/b_N = (D_{N}q^2+2\Delta_N^2+2\omega_{n}^2)^{-1}$. This sort of $q$-dependence signals that the interaction between impurities exponentially decreases at $r \gg \sqrt{D_N/\Delta_N}$. Unlike such an exponential dependence, the contribution associated with the condensate demonstrates a power-low behavior. Indeed, as can be seen from Eqs.(\ref{deltaS2}) and (\ref{deltaSN2}), at $q\rightarrow 0$ $\delta g_{N} \sim  \mathbf{qA}/q^2$. After the substitution of such  $\delta g_{N}$ into Eq.(\ref{sigma122}) and by taking into account Eq.(14) one arrives to the 2D dipole-dipole interaction given by Eq.(12).


\begin{thebibliography}{99}

\bibitem{Ruderman}
M. A. Ruderman and C. Kittel, Phys. Rev. \textbf{96}, 99 (1954).
\bibitem{Kasuya}
T. Kasuya, Progress of Theoretical Physics \textbf{16}, 45 (1956).
\bibitem{Yosida}
K. Yosida, Phys. Rev. \textbf{106}, 893 (1957).
\bibitem{Nadj}
S. Nadj-Perge, I. K. Drozdov, B. A. Bernevig, and A. Yazdani, Phys. Rev. B \textbf{88}, 020407 (2013).
\bibitem{Pientka}
F. Pientka,  L. I. Glazman, and Felix von Oppen, Phys. Rev. B \textbf{89}, 180505(R) (2014)
\bibitem{Christensen}
M. H. Christensen, M. Schecter, K. Flensberg, B. M. Andersen, and J. Paaske, Phys. Rev. B \textbf{94}, 144509 (2016)
\bibitem{Klinovaya}
J. Klinovaja, P. Stano, A. Yazdani, and D. Loss, Phys. Rev. Lett. \textbf{111}, 186805 (2013).
\bibitem{Menard}
G. C. M\'{e}nard, S. Guissart, C. Brun, R. T. Leriche, M. Trif, F. Debontridder, D. Demaille, D. Roditchev, P. Simon, and Tristan Cren, Nature Communications,  2017 DOI: 10.1038/s41467-017-02192-x
\bibitem{Pawlak}
Remy Pawlak,  Marcin Kisiel,  Jelena Klinovaja, Tobias Meier,  Shigeki Kawai,  Thilo Glatzel, Daniel Loss, and Ernst Meyer, npj Quantum Information \textbf{2}, 16035 (2016)
\bibitem{Li}
Jian Li, Titus Neupert, Zhi Jun Wang, A. H. MacDonald, A. Yazdani, and B. Andrei Bernevig, Nature Communications \textbf{7}, 12297 (2016)


\bibitem{Ptok}
 A. Ptok, Sz. Glodzik and T. Domanski,  Phys. Rev. B \textbf{96} 184425, (2017)


\bibitem{Hui}
Hoi-Yin Hui, P. M. R. Brydon, Jay D. Sau, S. Tewari1, and S. Das Sarma, Sci. Rep. \textbf{5}, 8880 (2015)
\bibitem{Nadj2}
S. Nadj-Perge, I. K. Drozdov, J. Li, H. Chen, S. Jeon, J. Seo, A. H. MacDonald, B. A. Bernevig, and A. Yazdani, Science \textbf{346}, 602 (2014).
\bibitem{Brydon}
P. M. R. Brydon, S. Das Sarma, H.-Y- Hui, and J. D. Sau, Phys. Rev. B\textbf{ 91}, 064505 (2015).
\bibitem{Heimes}
A. Heimes, D. Mendler, and P. Kotetes, New J. Phys. \textbf{17}, 023051 (2015).
\bibitem{Zyuzin}
A. A. Zyuzin and D. Loss, Phys. Rev. B \textbf{ 90}, 125443 (2014).
\bibitem{Qin}
Wei Quin and Zhenyu Zhang, Phys. Rev. Letters \textbf{113}, 266806 (2014)
\bibitem{Anderson}
P. W. Anderson and H. Suhl, Phys. Rev. \textbf{116}, 898 (1959)
\bibitem{Rashba}
Yu. A. Bychkov and E. I. Rashba, J. Phys. C \textbf{17}, 6039 (1984)
\bibitem{Yao}
N. Y. Yao, L. I. Glazman, E. A. Demler, M. D. Lukin, and J. D. Sau, Phys. Rev. Lett. \textbf{113}, 087202  (2014).
\bibitem{AGD}
A. A. Abrikosov, L. P. Gor'kov, and I. E. Dzyaloshinskii, Methods
of Quantum Field Theory in Statistical Physics (Dover, New York, 1975)
\bibitem{Nambu}
Y. Nambu, Phys. Rev. \textbf{117}, 648 (1960).
\bibitem{Eilenberger}
G. Eilenberger, Z.Phys. \textbf{214}, 195 (1968)
\bibitem{Larkin semiclass}
A. I. Larkin, and Y. N. Ovchinnikov, Zh. Eksp. Teor. Fiz. \textbf{55}, 2262 (1968) [Sov. Phys. JETP \textbf{28}, 1200 (1965)].
\bibitem{Rammer}
J. Rammer, H. Smith, Rev. Mod. Phys. \textbf{58}, 323 (1985)
\bibitem{Kopnin}
N. Kopnin, Theory of Nonequilibrium Superconductivity (Oxford Science, London, 2001).
\bibitem{Usadel}
K.D. Usadel, Phys. Rev. Lett. \textbf{25}, 507 (1970)
\bibitem{ZyuzinTI}
A. Zyuzin, M. Alidoust, and D. Loss, Phys. Rev. B \textbf{93}, 214502 (2016).
\bibitem{Bobkova}
I. V. Bobkova, A. M. Bobkov, A. A. Zyuzin, and M.Alidoust, Phys. Rev. B \textbf{94}, 134506 (2016)
\bibitem{Linder}
Henning G. Hugdal, Jacob Linder, and Sol H. Jacobsen, Phys. Rev. B 95, 235403 (2017)
\bibitem{bilayer}
A.G. Malshukov, Phys. Rev B. \textbf{95} , 064517 (2017)
\bibitem{Kupriyanov}
M. Yu. Kupriyanov and V. F. Lukichev, Zh. Eksp. Teor. Fiz.
\textbf{94}, 139 (1988) [Sov. Phys. JETP \textbf{67}, 1163 (1988)]
\bibitem{Zaitsev}
A. V. Zaitsev, JETP Lett. \textbf{51}, 35(1990)
\bibitem{Malsh island}
A.G. Mal'shukov, Phys. Rev. B \textbf{93}, 054511 (2016).
\bibitem{Pershoguba}
S. S. Pershoguba, K. Bj\"{o}rnson, A. M. Black-Schaffer, and
A. V. Balatsky, Phys. Rev. Lett. \textbf{115}, 116602 (2015).
\bibitem{Edelstein}
V. M. Edelstein, Sov. Phys. JETP \textbf{68}, 1244 (1989)
\bibitem{EdelsteinPRL}
V.M. Edelstein, Phys. Rev. Lett., \textbf{75}, 2004 (1996)
\end{thebibliography}
\end{document}